\documentclass[a4paper,11pt]{article}%
\usepackage{amsmath}
\usepackage{graphicx}
\usepackage{amsbsy}
\usepackage{cite}%
\usepackage{amsfonts}%
\usepackage{amssymb}

\newcommand{\bm}[1]{\boldsymbol{#1}}

\begin{document}
\thispagestyle{empty}  \setcounter{page}{0}
\begin{flushright}
May 2011 \\
\end{flushright}%

\vskip                                  3.8 true cm

\begin{center}
{\huge Proton stability from a fourth family}\\[2.15cm]

\textsc{Christopher Smith}\\[9pt]\textsl{Universit\'{e} Lyon 1 \& CNRS/IN2P3,
UMR5822 IPNL,}

\textsl{Rue Enrico Fermi 4, 69622 Villeurbanne Cedex, FRANCE}\\[2.2cm]

\textbf{Abstract}
\end{center}

\begin{quote}
\noindent The possibility to violate baryon or lepton number without
introducing any new flavor structures, beyond those needed to account for the
known fermion masses and mixings, is analyzed. With four generations, but only
three colors, this minimality requirement is shown to lead to baryon number
conservation, up to negligible dimension-18 operators. In a supersymmetric
context, this same minimality principle allows only superpotential terms with
an even number of flavored superfields, hence effectively enforces R-parity
both within the MSSM and in a GUT context. \let \thefootnote  \relax
\footnotetext{c.smith@ipnl.in2p3.fr} \newpage
\end{quote}

\section{Introduction}

The apparent stability of the proton is among the most puzzling phenomena, for
which a compelling theoretical explanation still eludes us. On one hand, its
lifetime is experimentally known to be greater than about $10^{30}$
years~\cite{PDG}. Even compared to the age of the Universe, this is overly
large. On the other hand, the conservation of the baryon ($\mathcal{B}$) and
lepton ($\mathcal{L}$) numbers is purely accidental in the Standard Model
(SM). It does not survive in many of its extensions, most notably in
supersymmetry. So, given the extremely long proton lifetime, naturality seems
to leave no other options than to forbid $\Delta\mathcal{B}$ and
$\Delta\mathcal{L}$ couplings entirely, either by hand in a phenomenological
context, or by a carefully designed theoretical setting in model-building
approaches~\cite{Report}.

The decay of the proton is usually presented as the archetype of the processes
induced by $\Delta\mathcal{B}$ and $\Delta\mathcal{L}$ interactions. But this
decay can also be regarded as a flavor transition, albeit of an extreme kind,
since some quark flavors transmute into lepton flavors. Said differently, the
$\Delta\mathcal{B}$ and $\Delta\mathcal{L}$ interactions are flavored
couplings. So, if they are not forbidden by some symmetries, it is their
flavor structures which have to be very special given the experimental
constraints from the proton lifetime and other $\Delta\mathcal{B}$ and
$\Delta\mathcal{L}$ observables. This observation suggests that the origin of
the proton stability could lie in the yet unknown mechanism from which all the
known flavor structures, i.e. the quark and lepton masses and mixings, derive.
The purpose of the present paper is to study whether this hypothesis is
tenable. As such, it departs from most model-building approaches, for which
the flavor structures of the $\Delta\mathcal{B}$ and $\Delta\mathcal{L}$
interactions are not central.

In principle, some information on the dynamics going on above the electroweak
scale is unavoidable to firmly establish a link between the flavor structures
of the interactions conserving and violating $\mathcal{B}$ and $\mathcal{L}$.
But, even without a full-fledged dynamical flavor theory, it is possible to
gather some information using low-energy effective theory techniques.
Specifically, these tools allow us to study how large $\Delta\mathcal{B}$ and
$\Delta\mathcal{L}$ interactions could be if they are required not to
introduce new flavor structures, beyond those already present in the SM. In
some sense, this provides a natural scale for the $\Delta\mathcal{B}$ and
$\Delta\mathcal{L}$ interactions, against which the experimental constraints
can be compared, as well as the predictions of a specific New Physics (NP) model.

This statement can be made more precise using the flavor-symmetry language.
From a low-energy perspective, all the flavored couplings break the
$U(N_{f})^{5}$ flavor symmetry of the SM gauge sector~\cite{Georgi}, with
$N_{f}$ the number of fermion flavors. The Yukawa couplings explicitly break
$SU(N_{f})^{5}$ but not $U(1)_{\mathcal{B},\mathcal{L}}\in U(N_{f})^{5}$,
while $\Delta\mathcal{B}$ and $\Delta\mathcal{L}$ interactions in general
break $U(N_{f})^{5}$ completely. A natural scale for the $\Delta\mathcal{B}$
and $\Delta\mathcal{L}$ interactions is obtained by preventing them from
introducing any new breaking of the $SU(N_{f})^{5}$ part of $U(N_{f})^{5}$.
Indeed, this symmetry requirement does not necessarily forbid them, but forces
them to be expressed entirely out of the Yukawa couplings, so that they
inherit the very peculiar hierarchies observed in the quark and lepton masses
and mixings. If these are sufficient to pass the experimental constraints, the
proton decay puzzle would somehow be alleviated since the required
fine-tunings of the $\Delta\mathcal{B}$ and $\Delta\mathcal{L}$ couplings
could then be viewed as reminiscent of those present in the known flavor
structures. Obviously, this would not yet be a solution to the proton decay
puzzle, since a dynamical model enforcing such a restricted flavor-breaking
sector is still missing, but it would nevertheless be a promising clue in that
direction. In this respect, our approach is similar in spirit to Minimal
Flavor Violation (MFV), see e.g.~Refs.~\cite{MFV,ComplexMFV,LeptonMFV}, but
here applied to $\Delta\mathcal{B}$ and $\Delta\mathcal{L}$
interactions~\cite{RPVMFV,LQMFV,Grossman}.

In this work, the number of flavors is central for a simple combinatorial
reason. As detailed in the following, the $\Delta\mathcal{B}$ or
$\Delta\mathcal{L}$ interactions necessarily involve epsilon contractions in
flavor space since they are written as invariant under $SU(N_{f})^{5}$, but
not under $U(N_{f})^{5}$. These epsilon tensors have as many indices as there
are generations. At the same time, Lorentz invariance requires an even number of fermion 
fields, and $\Delta\mathcal{B}=\pm1$ color singlets require three quark fields. 
So, going from three (odd) to four (even) generations completely changes the structure of the possible flavor-symmetric $\Delta\mathcal{B}$ interactions. These interactions systematically require  more quark fields,
and thus get suppressed, with four generations.

The present work focuses on the flavor properties of the $\Delta\mathcal{B}$
or $\Delta\mathcal{L}$ interactions, with or without an extra generation, but
not on the phenomenology of such a new generation. In this respect, it should
be noted that adding a generation of fermions is probably one of the simplest
extensions of the SM (see e.g. Ref.~\cite{4GReview} for reviews), and has
recently received quite some attention for example in the context of the
electroweak precision observables~\cite{4GEWO}, in connection to the possible
tensions exhibited in flavor physics~\cite{4GFlavor,4GFlavorB}, for the
discovery potential of colliders~\cite{4GCollider}, or for
baryogenesis~\cite{4GCosmo}. Experimentally, very recent searches for the
fourth-generation quarks have been performed at both Fermilab and at the LHC,
producing mass bounds from the direct $t^{\prime}$ and $b^{\prime}$ production
typically above 300-400 GeV~\cite{Searches} (though under some model-dependent
assumptions about their couplings to light quarks, see e.g.
Ref.~\cite{HiggsMix}). This means that fourth-generation Yukawa couplings must
be large, $m_{t^{\prime},b^{\prime}}/v\gtrsim2$, with $v\approx246$ GeV the SM
Higgs vacuum expectation value, and are thus pushed close to the threshold at
which perturbative calculations would cease to make sense.

The SM Higgs searches also tightly constrain the presence of a heavy fourth
generation, since it would enhance the Higgs production~\cite{SearchHiggs}
(see also Ref.~\cite{Gunion}). In this respect, the recent hint~\cite{CMSAtlas}
of a SM-like Higgs boson at around $126$ GeV is still inconclusive. Indeed,
the $\gamma\gamma$ signal is not much affected by a fourth generation, because
the enhancement of the $gg\rightarrow H$ production is mostly compensated by
the smaller $H\rightarrow\gamma\gamma$ rate \cite{Higgs126}, so better signals
in alternative channels are compulsory. Finally, it should be stressed that
the fourth generation discussed in the present work is purely sequential, i.e.
its fermions have exactly the same quantum numbers as those of the first three
generations, and get their masses from Yukawa couplings to the same SM Higgs
field. Alternative extended fermionic contents (see e.g.
Ref.~\cite{PFP,Perturb1}) or more complicated Higgs sectors (see e.g.
Refs.~\cite{Gunion,2HDM}) could be devised to ease some of the experimental
constraints, or preserve the perturbativity of the Lagrangian couplings.
Though we will not discuss these models, our approach based on the flavor
symmetry and its spurions is general enough, and could be easily adapted to
study the flavor structures of $\Delta\mathcal{B}$ or $\Delta\mathcal{L}$
interactions within these contexts.

The paper is organized according to the number and nature of the allowed
$SU(N_{f})^{5}$ breaking terms, since their restriction constitutes the
central working assumption used throughout this work. So, to systematically
study their impact, and compare the resulting $\Delta\mathcal{B}$ or
$\Delta\mathcal{L}$ operators in the three and four generation cases, we will
introduce these breaking terms gradually. In the next section, only the
$SU(3)_{C}\otimes SU(2)_{L}\otimes U(1)_{Y}$ gauge interactions are allowed,
so that $SU(N_{f})^{5}$ is unbroken. Then, in the following sections, the
complexity of the flavor-breaking sector is progressively increased by adding
the SM Higgs sector and its Yukawa couplings, the neutrino masses using a
seesaw model, the supersymmetric partners of SM particles, and finally, some
Grand Unified Theory (GUT) boundary conditions.

\section{$SU(3)_{C}\otimes SU(2)_{L}\otimes U(1)_{Y}$ gauge interactions}

The SM gauge interactions are identical for all the generations of matter
fields, denoted by the $Q=(u_{L},d_{L})$, $U=u_{R}^{\dagger}$, $D=d_{R}%
^{\dagger}$, $L=(\nu_{L},e_{L})$, and $E=e_{R}^{\dagger}$ left-handed Weyl
spinors. At the Lagrangian level, nothing distinguishes the $N_{f}$ flavors of
the five fermion species, and a large $U(N_{f})^{5}$ global flavor symmetry is
present. Since the $U(1)$s associated with $\mathcal{B}$ and $\mathcal{L}$ are
linear combinations of the five $U(1)$ factors of $U(N_{f})^{5}$, and since we
know that for such chiral symmetries, anomalies can arise, let us immediately
restrict the flavor symmetry to $G_{F}(N_{f})=SU(N_{f})^{5}$, and see what are
the possible $G_{F}(N_{f})$-symmetric interactions violating $\mathcal{B}$
and/or $\mathcal{L}$.

\paragraph{With three generations,}

only the $G_{F}(3)\equiv SU(3)^{5}$ symmetric contractions $\varepsilon
^{IJK}X^{I}X^{J}X^{K}$ with $X=Q,U,D,L,E$ have a nonzero $\mathcal{B}$ or
$\mathcal{L}$ charge ($I,J,K=1,2,3$ are generation indices; repeated indices
are summed over). An even number of such factors is needed to form a Lorentz
scalar. No $SU(3)_{C}\otimes SU(2)_{L}\otimes U(1)_{Y}$ invariant operators
violating $\mathcal{B}$ or $\mathcal{L}$ can be constructed with six fields,
so the leading $\Delta\mathcal{B}$ or $\Delta\mathcal{L}$ operators are
dimension 18:%
\begin{equation}
\mathcal{H}_{eff}^{gauge,SM3}=\frac{1}{\Lambda^{14}}((LQ^{3})^{3}%
+(EU^{2}D)^{3}+(EUQ^{\dagger2})^{3}+(LQD^{\dagger}U^{\dagger})^{3}%
+h.c.)\;,\label{gauge1}%
\end{equation}
where the flavor, $SU(2)_{L}$, $SU(3)_{C}$, and Lorentz spinorial contractions
are understood (only those contractions that maximally entwine the
antisymmetric tensors do not vanish identically). These operators are all
$(\Delta\mathcal{B},\Delta\mathcal{L})=\pm(3,3)$. Different patterns of
$\mathcal{B}$ and $\mathcal{L}$ violation are possible but require at least
six more fermion fields. For instance, with 18 fermion fields, dimension-27
operators inducing $\pm(\Delta\mathcal{B},\Delta\mathcal{L})=(6,0)$, $(0,6)$,
$(3,9)$, or $(3,\pm3)$ transitions can be written down.

At this level, in the absence of any flavor sector, it does not make much
sense to discuss the phenomenology or a possible NP origin for these
operators. Rather, the purpose of this construction is to point out a few
basic features of the procedure used throughout the paper.

First, the $G_{F}(3)$ symmetry requirement is very restrictive, since
$\Delta\mathcal{B}$ or $\Delta\mathcal{L}$ interactions are at least of
dimension 18, and cannot induce proton decay. By contrast, imposing only the
SM gauge invariance would allow for the dimension-six $LQ^{3}$, $EU^{2}D$,
$EUQ^{\dagger2}$, and $LQD^{\dagger}U^{\dagger}$ effective
interactions~\cite{Dim6}. We thus immediately conclude that these interactions
are not flavor-blind, i.e. require a nontrivial flavor structure to exist.
Second, among the operators in Eq.~(\ref{gauge1}), the one involving only
$SU(2)_{L}$ doublets, $(LQ^{3})^{3}$, can be recognized as arising from the
$U(1)_{\mathcal{B}+\mathcal{L}}$ anomaly~\cite{tHooft76}, i.e. from the
nonperturbative SM gauge dynamics. Though enforcing only the $G_{F}(3)$
symmetry requirement allows for additional operators in Eq.~(\ref{gauge1}), it is
nevertheless encouraging that it correctly predicts the order at which such
effects are generated. Further, the SM itself thus provides an example of how
the apparently peculiar $\varepsilon^{IJK}$ contractions in flavor space can
come into play. Third, inverting the argument, the anomalous nature of the
$U(3)^{5}$ symmetry justifies enforcing only $SU(3)^{5}$. Note, though, that
from a flavor point of view, the $(LQ^{3})^{3}$ operators violate
$U(1)_{Q}\otimes U(1)_{L}$ but respect $U(1)_{U}\otimes U(1)_{D}\otimes
U(1)_{E}$. So in principle, one can keep more than just $SU(3)^{5}$ as exact.

\paragraph{With four generations,}

the $\Delta\mathcal{L}$ or $\Delta\mathcal{B}$ contractions invariant under
$G_{F}(4)\equiv SU(4)^{5}$ must involve four fields, $\varepsilon^{IJKL}%
X^{I}X^{J}X^{K}X^{L}$. The crucial difference with the three generation case
is that for quarks, these monomials cannot be color singlets. Since the
smallest common multiple of three and four is twelve, three factors of such
quartic quark contractions are needed. The four-generation equivalents of the
operators in Eq.~(\ref{gauge1}) are thus of dimension 24:%
\begin{equation}
\mathcal{H}_{eff}^{gauge,SM4}=\frac{1}{\Lambda^{20}}((LQ^{3})^{4}%
+(EU^{2}D)^{4}+(EUQ^{\dagger2})^{4}+(LQD^{\dagger}U^{\dagger})^{4}+h.c.)\;,
\label{gauge2}%
\end{equation}
and induce $(\Delta\mathcal{B},\Delta\mathcal{L})=\pm(4,4)$. Again, the one
involving only $SU(2)_{L}$ doublets originates from the $\mathcal{B}%
+\mathcal{L}$ anomaly. However, another difference with the three generation
case is that here, the $G_{F}(4)$ symmetry is less restrictive, since
lower-dimensional $(\Delta\mathcal{B},\Delta\mathcal{L})=\pm(0,4)$ or
$(\Delta\mathcal{B},\Delta\mathcal{L})=\pm(4,0)$ operators can be constructed:%
\begin{equation}
\mathcal{H}_{eff}^{gauge,SM4}=\frac{1}{\Lambda^{14}}((LLE)^{4}+(UDD)^{4}%
+h.c.)\;. \label{gauge3}%
\end{equation}
This kind of couplings does not exist with three generations since they would
involve an odd number of fermion fields. So, for the SM with four generations,
the flavor symmetry fails at predicting the correct order at which
$\Delta\mathcal{B}$ and $\Delta\mathcal{L}$ effects can arise from the
nonperturbative SM gauge dynamics. Note, though, that if one imposes
$SU(4)^{5}\otimes U(1)_{U}\otimes U(1)_{D}\otimes U(1)_{E}$, the simplest
interaction is again, trivially, the anomalous one.

As for three generations, the current flavor-blind setting is too restrictive
to be relevant phenomenologically, and needs not be discussed further. Its
purpose was first to introduce the crucial difference between enforcing
$G_{F}(3)$ or $G_{F}(4)$ together with $SU(3)_{C}$, which stems from the
mismatch between the number of colors and flavors, and second, to show that
also with four generations, an operator can induce proton decay only if it has
a nontrivial flavor structure, i.e. a nontrivial behavior not only under the
$U(1)_{\mathcal{B},\mathcal{L}}$, but also under $G_{F}(4)$.

\begin{figure}[t]
\centering                        \includegraphics[width=16cm]{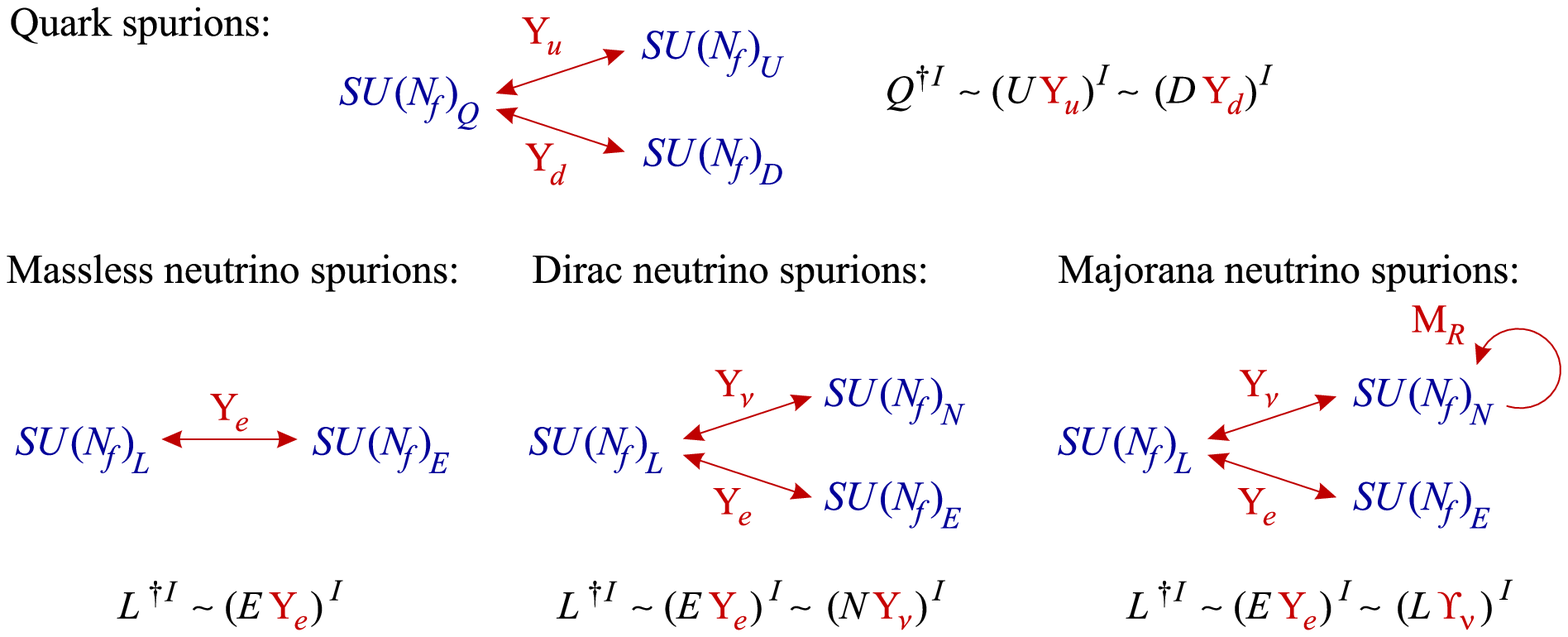}
\caption{Spurions in the quark and lepton sectors, as needed to induce the
known flavor structures. The Majorana mass term for the left-handed neutrino
is assumed to originate from a seesaw mechanism of type I, i.e. from a heavy
flavor-triplet of right-handed neutrinos. Those states are integrated out and
do not appear at low energy, see Eq.~(\ref{seesaw2}).}%
\label{Spurions}%
\end{figure}

\section{Adding the Higgs sector}

The Higgs sector breaks the $G_{F}(N_{f})$ flavor symmetry through the Yukawa
couplings%
\begin{equation}
\mathcal{L}_{\text{Yukawa}}=U\mathbf{Y}_{u}QH_{u}+D\mathbf{Y}_{d}%
QH_{d}+E\mathbf{Y}_{e}LH_{d}\;, \label{Yukawa}%
\end{equation}
since those involve fermions transforming according to different $SU(N_{f})$s.
Note that in the present section, neutrinos are still massless, and the Yukawa
couplings are written for a two-Higgs doublet model of type II in anticipation
of the Minimal Supersymmetric Standard Model (MSSM) discussion, but the SM case is trivially recovered through the
identifications $H_{u}\rightarrow H^{\ast}$, $H_{d}\rightarrow H$.

To systematically parametrize the impact of these breaking terms, the MFV
strategy is first to promote them to spurions, so as to artificially restore
the $G_{F}(N_{f})$ symmetry. Then, effective operators are constructed as
formally invariant under $G_{F}(N_{f})$, as well as under the gauge group
$SU(3)_{C}\otimes SU(2)_{L}\otimes U(1)_{Y}$, using the fermion fields and the
$\mathbf{Y}_{u,d,e}$ spurions as building blocks. Once done, the
$\mathbf{Y}_{u,d,e}$ spurions are frozen back to their physical values,
\begin{equation}
v_{u}\mathbf{Y}_{u}=\mathbf{m}_{u}V,\ \;v_{d}\mathbf{Y}_{d}=\mathbf{m}%
_{d},\;v_{d}\mathbf{Y}_{e}=\mathbf{m}_{e}\;, \label{background}%
\end{equation}
with $\mathbf{m}_{u,d,e}$ the $N_{f}\times N_{f}$ diagonal mass matrices, $V$
the $N_{f}\times N_{f}$ generalized CKM matrix, and $v_{u,d}$ the $H_{u,d}%
^{0}$ vacuum expectation values. In this way, the specific flavor-breaking
character of the Higgs sector, i.e. the transformation properties of its
spurions as well as their hierarchical structures, is exported onto the
effective operators.

Basically, the main feature of the Yukawa spurions is to interconnect the
flavor $SU(N_{f})$s, see Fig.~\ref{Spurions}. For example, $Q^{\dagger}$,
$U\mathbf{Y}_{u}$, and $D\mathbf{Y}_{d}$ all transform in the same way under
$SU(N_{f})_{Q}$. So, enforcing $G_{F}(N_{f})$ no longer means constructing
simple invariants like $\varepsilon^{IJK}Q^{I}Q^{J}Q^{K}$ for $N_{f}=3$ or
$\varepsilon^{IJKL}Q^{I}Q^{J}Q^{K}Q^{L}$ for $N_{f}=4$. This allows for
simpler effective operators, since monomials transforming identically under
$G_{F}(N_{f})$ need not have the same charges under the SM gauge group.

\paragraph{With three generations,}

and in the presence of the $\mathbf{Y}_{u,d,e}$ spurions, $\Delta\mathcal{B}$
or $\Delta\mathcal{L}$ operators arise much earlier than at $\mathcal{O}%
(\Lambda^{-14})$. Since three lepton or three quark fields are needed to form
a $\Delta\mathcal{L}\neq0$ or $\Delta\mathcal{B}\neq0$ but $G_{F}%
(3)$-invariant combination, the leading $SU(3)_{C}\otimes SU(2)_{L}\otimes
U(1)_{Y}$ invariant operators involve six fermion fields:%
\begin{equation}
\mathcal{H}_{eff}^{Yukawa,SM3}=\frac{1}{\Lambda^{5}}(EL^{\dagger2}%
U^{3}+L^{\dagger3}Q^{\dagger}U^{2}+D^{4}U^{2}+D^{3}UQ^{\dagger2}%
+D^{2}Q^{\dagger4}+h.c.)\;, \label{higgs1}%
\end{equation}
and respect either $(\Delta\mathcal{B},\Delta\mathcal{L})=\pm(1,3)$ or
$(\Delta\mathcal{B},\Delta\mathcal{L})=\pm(2,0)$. All these operators require
at least one $\mathbf{Y}_{u,d,e}$ insertion to form $G_{F}(3)$ singlets, for
example (spinor, $SU(2)_{L}$, and color contractions are understood)
\begin{subequations}
\label{higgs23}%
\begin{align}
L^{\dagger3}Q^{\dagger}U^{2}  &  =\varepsilon^{IJK}L^{\dagger I}L^{\dagger
J}L^{\dagger K}\otimes\varepsilon^{LMN}Q^{\dagger L}(U\mathbf{Y}_{u}%
)^{M}(U\mathbf{Y}_{u})^{N}+...\;,\;\;\label{higgs2}\\
EL^{\dagger2}U^{3}  &  =\varepsilon^{IJK}(E\mathbf{Y}_{e})^{I}L^{\dagger
J}L^{\dagger K}\otimes\varepsilon^{LMN}(U\mathbf{Y}_{u})^{L}(U\mathbf{Y}%
_{u})^{M}(U\mathbf{Y}_{u})^{N}+...\;. \label{higgs3}%
\end{align}
Also, none of them breaks $G_{F}(3)$ in the same way as Eq.~(\ref{gauge1}),
which remain the simplest operators satisfying $\Delta\mathcal{B}%
=\Delta\mathcal{L}$, even in the presence of the Yukawa spurions. In other
words, and without surprise, the anomalous breaking of $\mathcal{B}%
+\mathcal{L}$ does not spill over to lower dimensional operators. Still, it is
interesting to note that the flavor contractions in Eq.~(\ref{higgs23}) take
place entirely in the $SU(3)_{Q}\otimes SU(3)_{L}$ space, as for the anomalous
operator of Eq.~(\ref{gauge1}).

The $(\Delta\mathcal{B},\Delta\mathcal{L})=\pm(1,3)$ operators can induce
proton decay, but are very suppressed by the $G_{F}(3)$ symmetry, the limited
spurion content, and their high dimensionality. The former suppression comes
from the need to extract only first-generation up quarks, and no bottom quark
or tau lepton, while the epsilon antisymmetry asks for the flavors to be
different. To compensate, one needs to insert the nondiagonal $\mathbf{Y}%
_{u}$ as appropriate, and extract a $\nu_{\tau}$ instead of a $\tau$. From
Eq.~(\ref{higgs23}), the piece of the $L^{\dagger3}Q^{\dagger}U^{2}$ operator
contributing to proton decay ends up suppressed by $(m_{u}/v_{u})^{2}%
V_{ub}\sim10^{-13}$, while that from the $EL^{\dagger2}U^{3}$ operator by
$(m_{u}/v_{u})^{3}V_{us}V_{ub}\sim10^{-19}$. This flavor suppression, combined
with the overall factor of the order of $\mathcal{O}(m_{p^{+}}^{11}%
/\Lambda^{10})$ for the decay rate, ensures the proton lifetime is above about
$10^{30}$ years even for a relatively low NP scale, $\Lambda\gtrsim1$ TeV.
This is sufficient since the bounds on the $\Delta\mathcal{L}=\pm3$ decay
channels are much less tight than the best bound of $8.2\times10^{33}$
years~\cite{PDG} for the $\Delta\mathcal{L}=1$ mode $p^{+}\rightarrow e^{+}%
\pi^{0}$, which cannot be induced by~(\ref{higgs1}). These two suppression
mechanisms similarly ensure that the $\Delta\mathcal{B}=2$ operators do not
induce too rapid $n-\bar{n}$ oscillations.

So, in the presence of the Higgs sector, the simplest $\Delta\mathcal{B}$ and
$\Delta\mathcal{L}$ interactions that are naturally non vanishing are
dimension nine. They exhibit a strong hierarchy, inherited from the Yukawa
couplings, which suffices to pass all the current experimental bounds.

\paragraph{With four generations,}

the simplest $\Delta\mathcal{B}$ structures still require at least twelve
quark fields (least common multiple of four flavors and three colors), as for
the pure gauge situation. Proton decay as well as neutron oscillations are
thus forbidden. Said differently, in the presence of a fourth generation, an
operator inducing proton decay must have a flavor structure orthogonal to that
of the Yukawa couplings. If a NP model does not allow for this, then the
proton is stable.

By contrast, in the leptonic sector, $SU(3)_{C}\otimes SU(2)_{L}\otimes
U(1)_{Y}$ invariant operators with four lepton fields can be constructed,
\end{subequations}
\begin{equation}
\mathcal{H}_{eff}^{Yukawa,SM4}=\frac{1}{\Lambda^{6}}L^{4}H_{u}^{4}%
+\frac{1}{\Lambda^{7}}L^{4}H_{u}^{2}DU^{\dagger}+\mathcal{O}(\Lambda
^{-8})+h.c.\;, \label{higgs6}%
\end{equation}
which are $(\Delta\mathcal{B},\Delta\mathcal{L})=\pm(0,4)$. The first operator
being totally symmetric under the exchange of the four $SU(2)_{L}$
contractions $(LH_{u})$, a large number of $\mathbf{Y}_{e}$ insertions is
needed:%
\begin{equation}
L^{4}H_{u}^{4}=\varepsilon^{IJKL}(LH_{u})^{I}(\mathbf{Y}_{e}^{\dagger
}\mathbf{Y}_{e}LH_{u})^{J}((\mathbf{Y}_{e}^{\dagger}\mathbf{Y}_{e}%
\mathbf{)}^{2}LH_{u})^{K}((\mathbf{Y}_{e}^{\dagger}\mathbf{Y}_{e}%
\mathbf{)}^{3}LH_{u})^{L}+...\;, \label{higgs7}%
\end{equation}
resulting in a strong suppression by at least $m_{\mu}^{2}m_{\tau}^{4}%
m_{\tau^{\prime}}^{6}/v_{d}^{12}\sim\mathcal{O(}10^{-8})$ for $m_{\tau
^{\prime}}=500$ GeV and $\tan\beta=3$ (which is close to the perturbativity
limit with $m_{\tau^{\prime}}/v_{d}\approx5$). For that reason, the second
operator could be larger%
\begin{equation}
L^{4}H_{u}^{2}DU^{\dagger}=\varepsilon^{IJKL}(LH_{u})^{I}(\mathbf{Y}%
_{e}^{\dagger}\mathbf{Y}_{e}LH_{u})^{J}(L^{K}L^{L})(D\mathbf{Y}_{d}%
\mathbf{Y}_{u}^{\dagger}U^{\dagger})+... \label{higgs8}%
\end{equation}
With $J=4$, no light lepton mass factors are needed. Still, for $\Lambda
\gtrsim1$ TeV, the high-dimensionality together with the fact that one of the
external states is necessarily a heavy fourth generation lepton, $\nu
_{\tau^{\prime}}$ or $\tau^{\prime}$, makes this interaction difficult to
probe experimentally.

At this stage, it must be emphasized that these numerical estimates are valid
only in the perturbative regime. If the fourth generation is too heavy, the
MFV expansions are not predictive. Indeed, let us remind that the standard MFV
procedure relies not only on a restricted number of spurions, but also on a
strong naturality principle. Typically, MFV leads to polynomial expansions in
powers of the Yukawa couplings. For example, the expansion
\begin{equation}
a_{0}\mathbf{1}+a_{1}\mathbf{Y}_{e}^{\dagger}\mathbf{Y}_{e}+a_{2}%
(\mathbf{Y}_{e}^{\dagger}\mathbf{Y}_{e})^{2}+...\;\;\sim\;\;\mathbf{1}%
\oplus(\mathbf{N}_{f}\otimes\mathbf{\bar{N}}_{f})\;, \label{MFVexp}%
\end{equation}
transforming as the $SU(N_{f})_{L}$ adjoint can be inserted in several places
in Eqs.~(\ref{higgs7},~\ref{higgs8}). Importantly, all the numerical
coefficients (the $a_{i}$ of Eq.~(\ref{MFVexp})) are required to be natural,
i.e. of $\mathcal{O}(1)$. It is only in that case that the hierarchical
structures of the Yukawa couplings are passed on to the other flavor
couplings. But this is tenable only when the Yukawa couplings are also of
$\mathcal{O}(1)$ because for consistency, $G_{F}$-singlet traces like
$\langle\mathbf{Y}_{e}^{\dagger}\mathbf{Y}_{e}\rangle$ are implicitly absorbed
into these coefficients (this also follows from the Cayley-Hamilton theorem,
see Ref.~\cite{ComplexMFV}). So, if the Yukawa couplings ceased to be
perturbative, the naturality assumption collapses, the $a_{i}$ become
arbitrary, and MFV fails at transmitting the hierarchies of the Yukawa
couplings to the other flavored couplings. This would be particularly
problematic in the quark sector, since the MFV control over the $\Delta
\mathcal{B}=\Delta\mathcal{L}=0$ FCNC operators of relevance in $K$ and $B$
physics would be lost.

For our purpose, even if MFV may lose its original motivation drawn from the
tight experimental constraints on FCNC observables, this perturbativity issue
is only marginal. Indeed, the impossibility to control the coefficients
concerns only those operators allowed by the symmetry and spurion content.
But, with four generations, $\Delta\mathcal{B}=1$ operators are forbidden from
the start. As said, such an operator requires a new spurion to be allowed, and
this spurion must transform intrinsically differently than the Yukawa spurions
under $G_{F}(4)$.

A second issue is worth mentioning about the four-generation MFV
implementation discussed here. Even if perturbativity survives, the
$G_{F}(N_{f})$ version of MFV is much less predictive when $N_{f}>3$ since the
background values~(\ref{background}) are not entirely fixed. In the
four-generation case, the masses of the fourth-generation fermions as well as
the extended CKM and PMNS parameters are unknown. So, the FCNC operators
relevant for the tightly constrained $K$ and $B$ observables are not
completely controlled. This was discussed in Ref.~\cite{4GFlavorB}, where the
$G_{F}(4)$ version of MFV is projected onto a $G_{F}(3)$ structure by
factoring out a new $G_{F}(3)$ spurion accounting for the fourth-generation
degrees of freedom. The impact on the FCNC observables was then analyzed, as a
function of the assumed flavor structure of this new spurion.

For our purpose, this predictivity issue is not directly relevant. First, the
fourth generation fermions are not integrated out, but occur as dynamical
fields in the effective operators. The full $G_{F}(4)$ flavor symmetry of the
gauge sector is active, and there is no need to project onto $G_{F}(3)$.
Second, even if the unknown $G_{F}(4)$ spurion parameters render numerical
estimates uncertain, like for the operators in Eqs.~(\ref{higgs7}%
,~\ref{higgs8}), they do not play any role in the construction of the
$\Delta\mathcal{B}$ or $\Delta\mathcal{L}$ effective operators, which is
entirely fixed by the $G_{F}(4)$ symmetry properties of the spurions.

\section{Adding neutrino masses}

In the previous section, neutrinos were left massless, as in the minimal
version of the Standard Model. The simplest way to give them a mass is to add
the Yukawa coupling%
\begin{equation}
\mathcal{L}^{\text{Dirac}}=N\mathbf{Y}_{\nu}LH_{u}\;,
\end{equation}
to Eq.~(\ref{Yukawa}), with $N=\nu_{R}^{\dagger}$ a flavor $N_{f}$-plet of
right-handed neutrinos (see Fig.~\ref{Spurions}). However, neither the
presence of $\mathbf{Y}_{\nu}$ among the spurions nor the possibility to write
operators involving $N$ fields change the previous conclusions. Obviously,
$\Delta\mathcal{B}$ operators are unaffected by the spurion content of the
leptonic sector. Further, $\mathcal{L}$ can still change only in step of
$N_{f}$, so the simplest operators have the same $\mathcal{B}$ and
$\mathcal{L}$ overall charges as in Eq.~(\ref{higgs1}) or~(\ref{higgs6}), but
for the new $(\Delta\mathcal{B},\Delta\mathcal{L})=\pm(0,6)$ operator $N^{6}$
permitted in the three generation case.

The situation changes if left-handed neutrinos have a Majorana mass term, as
arising from the $(\Delta\mathcal{B},\Delta\mathcal{L})=\pm(0,2)$
dimension-five Weinberg operator:%
\begin{equation}
\mathcal{H}_{eff}^{\text{Majorana}}=(LH_{u})^{T}\frac{\mathbf{m}_{\nu}}%
{v_{u}^{2}}(LH_{u})\;. \label{seesaw1}%
\end{equation}
Such an effective interaction typically arises from the type-I seesaw
mechanism~\cite{Seesaw}, which relates $\mathbf{m}_{\nu}$ to the right-handed
neutrino Majorana mass $\mathbf{M}_{R}$ and neutrino Yukawa couplings
$\mathbf{Y}_{\nu}$ as%
\begin{equation}
\mathbf{\Upsilon}_{\nu}^{\dagger}\equiv\frac{\mathbf{m}_{\nu}}{v_{\nu}}%
=v_{u}\mathbf{Y}_{\nu}^{T}\mathbf{M}_{R}^{-1}\mathbf{Y}_{\nu}\;.
\label{seesaw2}%
\end{equation}
(the $\dagger$ is introduced only for future convenience.) In this way, one
avoids introducing unnaturally small Lagrangian couplings in the neutrino
sector since $\mathbf{Y}_{\nu}$ can be of $\mathcal{O}(1)$ when $\mathbf{M}%
_{R}\approx10^{13}$ GeV.

For simplicity, in the present analysis, we add only the effective
$\mathbf{\Upsilon}_{\nu}$ spurion to $\mathbf{Y}_{u,d,e}$. This spurion is
symmetric, transforms as $\mathbf{N}_{f}\otimes\mathbf{N}_{f}$ under
$SU(N_{f})_{L}$, but has tiny background values, with entries at most of
$\mathcal{O}(10^{-12})$ for the first three generations. On the other hand,
the spurion combination $\mathbf{Y}_{\nu}^{\dagger}\mathbf{Y}_{\nu}$,
characteristic of an underlying seesaw mechanism, is not
suppressed\footnote{This spurion combination is mostly relevant for lepton
flavor violating effects, since $\mathbf{Y}_{\nu}^{\dagger}%
\mathbf{Y}_{\nu}$ and $\mathbf{Y}_{e}^{\dagger}\mathbf{Y}_{e}$ are not
simultaneously diagonal. The current experimental bounds on the lepton
flavor violating transitions like $\ell^{I}\rightarrow\ell^{J}\gamma$, $I\neq J$, severely
limit these off-diagonal entries for NP scales at or below the TeV scale,
see e.g. Ref. \cite{LeptonMFV}.} but irrelevant for the construction of
$\Delta\mathcal{L}$ operators since it transforms as $\mathbf{Y}_{e}^{\dagger
}\mathbf{Y}_{e}$, and will thus be left out.

\paragraph{With three generations,}

there is no more need to go fetch the $\Delta\mathcal{L}=3$ terms, as in
Eq.~(\ref{higgs1}), because $\Delta\mathcal{L}=1$ interactions are allowed
through the generic contraction $\varepsilon^{IJK}(\mathbf{\Upsilon}_{\nu
}\mathbf{Y}_{e}^{\dagger}\mathbf{Y}_{e})^{IJ}L^{K}$ (the $\mathbf{Y}%
_{e}^{\dagger}\mathbf{Y}_{e}$ insertion is needed to compensate for the
symmetry\footnote{There is no singlet in $\mathbf{3}\otimes\mathbf{6}%
=\mathbf{10}\oplus\mathbf{8}$, but $\mathbf{Y}_{e}^{\dagger}\mathbf{Y}_{e}%
\sim\mathbf{8}$ permits to extract the one occurring in $\mathbf{3}%
\otimes(\mathbf{6}\otimes\mathbf{8)}.$} of $\mathbf{\Upsilon}_{\nu}$). So, the
simplest interactions are the dimension-six $(\Delta\mathcal{B},\Delta
\mathcal{L})=\pm(1,1)$ Weinberg operators~\cite{Dim6}%
\begin{equation}
\mathcal{H}_{eff}^{Seesaw,SM3}=\frac{1}{\Lambda^{2}}(LQ^{3}+EU^{2}%
D+EUQ^{\dagger2}+LQD^{\dagger}U^{\dagger}+h.c.)\;. \label{seesaw3}%
\end{equation}
These operators cannot be trivially discarded since $\mathcal{B}+\mathcal{L}$
is not a good symmetry. In other words, they have the same $U(1)$ charges
(modulo three) as the operators~(\ref{gauge1}). Including the Higgs fields,
$(\Delta\mathcal{B},\Delta\mathcal{L})=\pm(1,-1)$ dimension-seven operators
can be constructed (still proportional to $\mathbf{\Upsilon}_{\nu}$), along
with corrections to the Majorana mass term for the neutrinos.

It is only once neutrinos get a Majorana mass term that MFV permits to set a
nontrivial natural scale for the Wilson coefficients of the Weinberg
operators. But, neutrinos being so light, these couplings will automatically
be very suppressed. A further suppression comes from the epsilon contractions,
exactly like for the operators in Eq.~(\ref{higgs23}), since their
antisymmetry asks for the flavors of the quark fields to be all different,
while only $u$, $d$, $s$ can be present to induce proton decay. In the
Appendix, we explicitly show that numerically, these two mechanisms are
sufficient to pass the proton decay bounds even for a relatively low NP scale,
$\Lambda\gtrsim10$ TeV. As discussed there, it may even be possible to pass
these bounds at the electroweak scale, provided Yukawa coupling insertions are
accompanied by gauge couplings and loop factors.

\paragraph{With four generations,}

the situation is again very different. The presence of $\mathbf{\Upsilon}%
_{\nu}$ in the leptonic sector does not change anything for $\Delta
\mathcal{B}$ operators, which still need twelve quark fields, as in
Eqs.~(\ref{gauge2},~\ref{gauge3}). Thus, proton decay and neutron oscillations
remain forbidden.

In the leptonic sector, $\mathbf{\Upsilon}_{\nu}$ brings in an even number of
$SU(4)_{L}$ flavor indices, so its main impact is to allow for new
$\Delta\mathcal{L}=\pm2$ operators (operators with $H_{u}\leftrightarrow
H_{d}^{\dagger}$ are understood):%
\begin{align}
\mathcal{H}_{eff}^{Seesaw,SM4}  &  =\frac{\varepsilon^{IJKL}}{\Lambda
}(\mathbf{\Upsilon}_{\nu})^{IJ}(LH_{u})^{K}(LH_{u})^{L}+\frac{\varepsilon
^{IJKL}}{\Lambda^{3}}(\mathbf{\Upsilon}_{\nu})^{IJ}(LH_{u})^{K}(LH_{u}%
)^{L}H_{u}H_{d}\label{seesaw4}\\
&  +\frac{\varepsilon^{IJKL}}{\Lambda^{3}}(\mathbf{\Upsilon}_{\nu})^{IJ}%
L^{K}L^{L}H_{u}(D\mathbf{Y}_{d}Q+E\mathbf{Y}_{e}L+Q^{\dagger}\mathbf{Y}%
_{d}^{\dagger}U^{\dagger})+\mathcal{O}(\Lambda^{-5})+h.c.\;.
\end{align}
The dimension-five operator contributes to the Majorana mass term itself. But
since it requires some Yukawa insertions to be nonzero ($\mathbf{\Upsilon
}_{\nu}$ and $(LH_{u})^{2}$ are symmetric in flavor space), the induced effect
on $\mathbf{m}_{\nu}$ is subleading. The operators in the second line
contribute to $\Delta\mathcal{L}=\pm2$ flavor transitions like for example
$K^{+}\rightarrow\pi^{-}\ell^{+}\ell^{+}$, but are presumably too suppressed
by the several factors of the light fermion mass to be accessible
experimentally, even for large fourth-generation entries in $\mathbf{\Upsilon
}_{\nu}$~\cite{Majorana}. In this respect, let us stress that the same
provision on the perturbativity of the spurions as in the previous section is
required for MFV expansions to be valid. As before, this is irrelevant for
proton decay, since no expansion can be written for $\Delta\mathcal{B}=\pm1$
couplings, given the assumed spurion content.

\section{Adding supersymmetric partners}

Up to now, an even number of matter fields was required to form Lorentz invariant. This no longer holds when those fields receive scalar superpartners, with which they share their quantum numbers. The consequence is well known: the MSSM a priori allows for the renormalizable $\Delta\mathcal{B}$ or $\Delta\mathcal{L}$ interactions~\cite{Barbier04}%
\begin{equation}
\mathcal{W}_{\Delta\mathcal{B},\Delta\mathcal{L}}=\frac{1}{2}\boldsymbol{\lambda}^{IJK}L^{I}L^{J}%
E^{K}+\boldsymbol{\lambda}  ^{\prime IJK}L^{I}Q^{J}D^{K}+\boldsymbol{\mu}
^{\prime I}H_{u}L^{I}+\frac{1}{2}\boldsymbol{\lambda}  ^{\prime\prime
IJK}U^{I}D^{J}D^{K}\;, \label{susy1}%
\end{equation}
where $Q,U,D,L,E$ now denote superfields (the corresponding soft-breaking terms are understood throughout this section). With these couplings of order one, the proton would decay very quickly. So, phenomenologically, they are usually discarded by enforcing by hand a new discrete symmetry, R-parity~\cite{FarrarF78}.

The R-Parity Violating (RPV) couplings are flavored: they not only break
$U(1)_{\mathcal{B}}$ and $U(1)_{\mathcal{L}}$, but also $SU(N_{f})^{5}$. So,
following the philosophy exposed in the Introduction, a natural size for these
couplings can be derived by forbidding them from introducing any new flavor
structure. In other words, their natural size is obtained by expressing them
as $G_{F}(N_{f})$ invariant combinations of the $\mathbf{Y}_{u,d,e}$ and
$\mathbf{\Upsilon}_{\nu}$ spurions. The main phenomenological consequence can
immediately be guessed from the previous sections. Indeed, integrating out the
sparticles, the RPV couplings generate precisely the effective $\Delta
\mathcal{B}$ and/or $\Delta\mathcal{L}$ operators studied before. Since in
addition the same minimal spurion content is assumed, we immediately know that
the proton decays sufficiently slowly in the three-generation case, and is
absolutely stable in the four-generation case, where only $\Delta
\mathcal{B}=\pm4$ operators can be constructed.

At this stage, one may be a bit puzzled by the asymmetric status given to the
R-parity conserving and violating couplings. After all, they are all
renormalizable couplings of the MSSM superpotential\footnote{Further, they get
mixed by the possible redefinitions needed to get physical fields out of the
Lagrangian $H_{d}$ and $L$ fields. The consistency of the MFV expansion in the
presence of this freedom was analyzed in details in Ref.~\cite{LQMFV}.}. The
procedure we follow can be interpreted in several ways:

\begin{itemize}
\item From a purely low-energy perspective, there is no reason to expect the
RPV couplings to be flavor blind (e.g., $\boldsymbol{\lambda} ^{IJK}%
=\boldsymbol{\lambda} $ for all $I,J,K$), especially given the highly
hierarchal SM flavor structures. So, writing them down in terms of the known
flavor structures, $\mathbf{Y}_{u,d,e}$ and $\mathbf{\Upsilon}_{\nu}$, is an
attempt at assigning to them a similarly nonuniversal flavor structure in a
controlled way.

\item Taking the opposite view, it is obvious that assigning by hand a sufficiently peculiar $SU(N_{f})^{5}$ flavor structure to each RPV coupling can suppress the proton decay to an acceptable rate. However, it is generally believed that such assignments are too fine-tuned to be acceptable, and consequently, that only an exact symmetry can prevent rapid proton decay. The present analysis shows that the fine-tunings of the RPV couplings are in reality no less natural (or less unnatural) than the strong hierarchies exhibited by the known quark and lepton masses and mixings.

\item From a high-energy perspective, it is reasonable to expect that none of
the low-energy flavor couplings is fundamental. But, if they all derive from a
limited set of fundamental high-energy spurions, they should be related since
they are ultimately redundant. Though a full-fledged dynamical theory of
flavor would be required to consistently implement this picture, the MFV
procedure can be seen as an attempt at capturing such relationships. So, under
the assumption that the low-energy flavor couplings are (maximally) redundant,
expressing the unknown RPV couplings in terms of the known quark and lepton
masses and mixings does not presume anything about their relative fundamentality.

\item Finally, one can also consider the present section as a kind of
consistency check for the previous, purely effective discussions. It
illustrates how enforcing MFV on a well-known NP model with $\Delta
\mathcal{B}$ or $\Delta\mathcal{L}$ interactions leads to the suppression
obtained in a model-independent way.
\end{itemize}

To a large extent, one can recognize in the above points an exact translation
in the MSSM context of the general strategy presented in the Introduction.
These points were also discussed in Ref.~\cite{RPVMFV}, dedicated exclusively
to the consequences of MFV for the RPV couplings.

Let us now particularize the discussion to the three and four generation case,
and examine the phenomenological consequences.

\paragraph{With three generations,}

the minimal spurion content allows for all the RPV couplings. This
construction was performed in detail in Ref.~\cite{RPVMFV}, to which we refer
for more information.

Briefly, the $(\Delta\mathcal{B},\Delta\mathcal{L})=\pm(1,0)$ RPV coupling is
expressible entirely out of $\mathbf{Y}_{u,d}$, for example as%
\begin{align}%
\bm{\lambda}%
_{MSSM3}^{\prime\prime IJK}  &  =\lambda_{1}^{\prime\prime}\varepsilon
^{LJK}(\mathbf{Y}_{u}\mathbf{Y}_{d}^{\dagger})^{IL}+\lambda_{2}^{\prime\prime
}\varepsilon^{IMN}(\mathbf{Y}_{d}\mathbf{Y}_{u}^{\dagger})^{JM}(\mathbf{Y}%
_{d}\mathbf{Y}_{u}^{\dagger})^{KN}+\lambda_{3}^{\prime\prime}\varepsilon
^{LMN}\mathbf{Y}_{u}^{IL}\mathbf{Y}_{d}^{JM}\mathbf{Y}_{d}^{KN}\nonumber\\
&  +\lambda_{4}^{\prime\prime}\varepsilon^{LMN}\varepsilon^{IAB}%
\varepsilon^{DJK}\mathbf{Y}_{d}^{\dagger LD}\mathbf{Y}_{u}^{\dagger
MA}\mathbf{Y}_{u}^{\dagger NB}+...\;,
\end{align}
with $\lambda_{i}^{\prime\prime}\sim\mathcal{O}(1)$. So,
$\boldsymbol{\lambda}  ^{\prime\prime}$ can be large, up to $\mathcal{O}(1)$
values for $%
\bm{\lambda}%
^{\prime\prime312}$ when $\tan\beta=v_{u}/v_{d}\gtrsim20$, but also show a
strong hierarchy. On the other hand, the leptonic Yukawa couplings
$\mathbf{Y}_{e}$ and $\mathbf{Y}_{\nu}$ permit to change $\mathcal{L}$ only in
step of three. So, the $(\Delta\mathcal{B},\Delta\mathcal{L})=\pm(0,1)$ RPV
couplings require the presence of a Majorana mass term for the neutrinos, for
example as%
\begin{align}%
\bm{\mu}%
_{MSSM3}^{\prime I}  &  =\mu_{1}^{\prime}\varepsilon^{IJK}(\mathbf{\Upsilon
}_{\nu}\mathbf{Y}_{e}^{\dagger}\mathbf{Y}_{e})^{JK}+...\;,\\%
\bm{\lambda}%
_{MSSM3}^{IJK}  &  =\lambda_{1}\varepsilon^{ILM}(\mathbf{\Upsilon}_{\nu
}\mathbf{Y}_{e}^{\dagger}\mathbf{Y}_{e})^{LM}\mathbf{Y}_{e}^{KJ}+\lambda
_{2}\varepsilon^{IMJ}(\mathbf{Y}_{e}\mathbf{\Upsilon}_{\nu})^{KM}+...\;,\\%
\bm{\lambda}%
_{MSSM3}^{\prime IJK}  &  =\lambda_{1}^{\prime}\varepsilon^{ILM}%
(\mathbf{\Upsilon}_{\nu}\mathbf{Y}_{e}^{\dagger}\mathbf{Y}_{e})^{LM}%
\mathbf{Y}_{d}^{KJ}+...\;,
\end{align}
with $\mu_{i}$, $\lambda_{i}$, and $\lambda_{i}^{\prime}$ of $\mathcal{O}(1)$.
Note that $\mathbf{Y}_{e}^{\dagger}\mathbf{Y}_{e}$ factors have been
introduced to compensate for the symmetry of $\mathbf{\Upsilon}_{\nu}$. So,
these RPV couplings are all suppressed by the tiny neutrino masses, as well as
by some charged lepton masses.

The suppression of the RPV couplings brought in by the proportionality to
$\mathbf{\Upsilon}_{\nu}$ and by the antisymmetry of the epsilon contractions
is sufficient to pass the bounds on proton decay (as well as on all other
$\Delta\mathcal{B}$ or $\Delta\mathcal{L}$ observables, see Ref.~\cite{RPVMFV}%
). This can be understood from the numerical analysis of the Weinberg
operators presented in the Appendix. Indeed, integrating out the squarks and
slepton, the only effective operator accessible from the couplings of
Eq.~(\ref{susy1}) is $L^{I}Q^{J}D^{\dagger L}U^{\dagger K}$, proportional to
$\boldsymbol{\lambda} ^{\prime IJM}\boldsymbol{\lambda} ^{\prime\prime KLM}$.
Under MFV, this operator ends up much more significantly suppressed than
$LQ^{3}$ because light-quark mass factors necessarily occur (see
Eq.~(\ref{C2IJKL})). So, the scale $\Lambda$, now identified with the $D^{M}$
mass, can be of a few hundred GeV. The only new feature compared to the
Appendix is that the predicted rates show a strong dependence on $\tan\beta$.
If it is large, the current proton decay bounds require relatively heavy
squarks, of at least a few TeV~\cite{RPVMFV}. Note, finally, that if a
holomorphic constraint is imposed on the spurions, as proposed recently in
Ref.~\cite{Grossman}, even this operator becomes forbidden, and the leading
proton decay mechanism is induced only at the loop level by to the
$\boldsymbol{\lambda} ^{IJK}$ and $\boldsymbol{\lambda} ^{\prime\prime IJK}$
couplings. Current experimental bounds are then easily satisfied, even for
light sparticles.

In addition to the superpotential terms~(\ref{susy1}), nonrenormalizable
interactions may be present since the MSSM is most probably only an effective
low-energy theory. Those have the same forms as the effective operators
discussed in the previous sections, up to the replacement of SM fermionic
fields by their corresponding superfields, and the flavor symmetry
requirements are identical. Consider for example the dimension-four
superpotential terms of the form~(\ref{seesaw3}) which can induce proton
decay~\cite{IbanezR91}. Those actually conserve R-parity, so it is only at the
loop level, with the help of gauge interactions, that they contribute to
proton decay. With these gauge couplings and loop factors, the proportionality
to $\mathbf{\Upsilon}_{\nu}$, and the antisymmetric contractions, proton decay
bounds are easily satisfied, even for relatively low NP scales~\cite{RPVMFV}.

\paragraph{With four generations,}

none of the $\mathcal{W}_{\Delta\mathcal{B},\Delta\mathcal{L}}$ terms are allowed,%
\begin{equation}
\boldsymbol{\mu}  _{MSSM4}^{\prime I}=\boldsymbol{\lambda}  _{MSSM4}%
^{IJK}=\boldsymbol{\lambda}  _{MSSM4}^{\prime IJK}=\boldsymbol{\lambda}
_{MSSM4}^{\prime\prime IJK}=0\;,
\end{equation}
simply because there is no way to contract an odd number of flavor indices
using the available spurions (two indices) and the epsilon tensors (four
indices). In other words, the transformation properties of the RPV couplings
under $SU(4)^{5}$ are orthogonal to those of the known flavor structures. So,
the minimal spurion content effectively enforces R-parity, the proton decay is
forbidden, and the lightest supersymmetric particle is stable.

The consequences for higher dimensional operators, if present, can be readily
drawn from the previous sections. The $\Delta\mathcal{B}$ couplings require no
less than twelve quark superfields, so $\mathcal{B}$ is effectively conserved.
Also at the effective level, neither proton decay nor neutron oscillations are
permitted. On the other hand, $\mathcal{L}$ can be violated, but exclusively
by an even number. So, R-parity emerges naturally also for nonrenormalizable
operators, making it a good effective symmetry.

In addition, the $\Delta\mathcal{L}=2$ operators are suppressed by neutrino
masses, so the dominant $\Delta\mathcal{L}$ effects arise from the superfield
version of the $\Delta\mathcal{L}=\pm4$ operators of Eq.~(\ref{higgs6}). As
discussed there, those are significantly suppressed by the NP scale, require
at least four leptons, and thus should have no impact on low-energy phenomenology.

\section{Adding GUT boundary conditions}

A crucial feature of the spurions is their separation into two sectors,
leptonic and baryonic, see Fig.~\ref{Spurions}. The absence of connection
between the $SU(N_{f})_{Q}\otimes SU(N_{f})_{U}\otimes SU(N_{f})_{D}$ and the
$SU(N_{f})_{L}\otimes SU(N_{f})_{E}$ flavor spaces effectively factorizes the
conservation or nonconservation of $\mathcal{B}$ and $\mathcal{L}$. For
instance, with three generations, $\Delta\mathcal{B}=\mathbb{Z}$ and
$\Delta\mathcal{L}=3\mathbb{Z}$ interactions are allowed by the Yukawa
couplings, but $\Delta\mathcal{L}=\mathbb{Z}$ interactions are proportional to
the neutrino Majorana masses. With four generations, only $\Delta
\mathcal{B}=4\mathbb{Z}$ and $\Delta\mathcal{L}=2\mathbb{Z}$ are allowed
($\Delta\mathcal{L}=4\mathbb{Z}$ without neutrino Majorana masses). These
``selection rules'', when combined with the $SU(3)_{C}\otimes SU(2)_{L}\otimes
U(1)_{Y}$ invariance, strongly constrain the Wilson coefficients of
$\Delta\mathcal{B}$, $\Delta\mathcal{L}$ effective operators, or the RPV
couplings of the MSSM.

In the present section, the goal is to illustrate what could occur if such a
factorization does not survive beyond the SM or MSSM. To connect these two
sectors realistically, GUTs like $SU(5)$ or $SO(10)$
appear tantalizing, especially given our focus on proton decay. At the same
time, the GUT context with its rather precise dynamics and flavor structure
represents a challenge for MFV, both at the conceptual and practical level.
Let us comment on four particular issues:

\begin{enumerate}
\item \textbf{Generic MFV implementation in GUT}: The flavor group above the
unification scale is smaller than $SU(N_{f})^{5}$, since there are less
fermion species~\cite{MFVSU5}. For example, in $SU(5)$, there are two fermion
multiplets per generation, $\mathbf{\bar{5}}$ and $\mathbf{10}$, so the flavor
group is $SU(N_{f})_{\bar{5}}\otimes SU(N_{f})_{10}$. Similarly, in $SO(10)$,
each generation of fermions is put in a $\mathbf{16}$ representation, and the
flavor group is simply $SU(N_{f})_{16}$. Further, the spurion structure does
not match the SM one. For example, in $SU(5)$, the spurions $\mathbf{Y}%
_{10}\sim(\mathbf{1},\mathbf{N}_{f}\otimes\mathbf{N}_{f})$ and $\mathbf{Y}%
_{5}\sim(\mathbf{\bar{N}}_{f},\mathbf{N}_{f})$ are not bifundamental
representations of the SM flavor group $SU(N_{f})^{5}$. As a result, matching
an $SU(N_{f})_{\bar{5}}\otimes SU(N_{f})_{10}$ MFV expansion onto an
$SU(N_{f})^{5}$ expansion, the quark Yukawa couplings end up in the leptonic
sector, and vice versa. Further, naturality may be lost in this projection
because the coefficients of the $SU(N_{f})^{5}$ expansion are not of
$\mathcal{O}(1)$ in general. Even though the RGE tend to bring generic
expansions back to MFV~\cite{ComplexMFV}, whether this is always sufficient to
recover naturality at low energy is not yet established (especially for
$N_{f}\neq3$, see point 3 below). Finally, to make matters worse, the fermion
mass unification is delicate, and requires extending the minimal GUT spurion
content. Besides the increased complexity of the MFV expansions, their
predictivity is seriously affected because the background values of these
spurions are not entirely fixed~\cite{MFVSU5}.

\item \textbf{Proton decay through GUT gauge interactions}: In GUT, baryon and
lepton numbers are no longer good quantum numbers since quarks and leptons are
unified. Said differently, $U(1)_{\mathcal{B}}$ and $U(1)_{\mathcal{L}}$ are
not contained in $U(N_{f})_{\bar{5}}\otimes U(N_{f})_{10}$ or $U(N_{f})_{16}$.
Further, the gauge interactions induce proton decay through leptoquark
tree-level exchanges. In the MFV language, the leptoquark gauge couplings
become, after the spontaneous breaking down to the $SU(3)_{C}\otimes
SU(2)_{L}\otimes U(1)_{Y}$ gauge group, genuine new $SU(N_{f})^{5}$ spurions
connecting the leptonic and hadronic flavor spaces. Those allow
the dimension-six $(\Delta\mathcal{B},\Delta\mathcal{L})=\pm(1,1)$ effective
operators of Eq.~(\ref{seesaw3}), with the scale $\Lambda=M_{GUT}$ set by the
leptoquark gauge boson masses. MFV has no handle on these spurions, since they
are protected by the GUT gauge symmetry, and thus cannot be assumed related to
the Yukawa spurions in any way. So, one has to sufficiently increase the
leptoquark masses, i.e. the unification scale. This does not invalidate the
MFV framework per se, since the high unification scale required by proton
decay is in any case (roughly) compatible with that obtained from the gauge
coupling unification, but it shows its limitation. When proton decay is
induced by the GUT gauge interactions, MFV is entirely irrelevant.

\item \textbf{Perturbativity up to the GUT scale}: As said, proton decay
requires $M_{GUT}$ to be rather close to the Planck scale. At the same time,
if $N_{f}>3$, the Yukawa couplings of the last $N_{f}-3$ generations have to
be close to the perturbativity limit already at the electroweak scale, and
certainly do not remain perturbative up to the GUT
scale~\cite{SU5,Perturb1,Perturb0}. This is particularly serious in a
supersymmetric setting, since the Higgs mass requires $\tan\beta$ not too
small but the current bound on $m_{b^{\prime}}$ is already above the
electroweak scale~\cite{PDG}. Though it may actually become a welcome feature
in the context of the electroweak symmetry breaking~\cite{EW}, the onset of a
nonperturbative regime not far from the TeV scale may render the GUT setting
difficult to manage in practice. If one insists on perturbativity, one way to
proceed is to alter the particle content, so as to modify the evolution up to
the GUT scale~\cite{Perturb1}. But then, to have an impact, the extra
particles should couple to quarks and leptons, i.e. should introduce new
flavor couplings. Whether these couplings have to be part of the spurions or
can be constructed out of the Yukawa spurions, the consequences for the MFV
expansions, for proton decay, and for FCNC constraints, have to be analyzed on
a case-by-case basis.

\item \textbf{Supersymmetric GUT and R-parity}: Even if the proton decay
mechanism through gauge interactions is controlled by pushing the unification
scale close to the Planck scale, this may not suffice. Indeed, in a
supersymmetric setting, a resurgence of the RPV couplings~(\ref{susy1}) is
possible, either directly from specific GUT-scale couplings, or concurrently
with the GUT spontaneous breaking chain. For example, in $SU(5)$, the
following trilinear coupling can be constructed
\begin{equation}
\mathcal{W}_{RPV}=\boldsymbol{\lambda}  _{5}^{IJK}\bar{5}^{I}\bar{5}^{J}%
10^{K}\;, \label{RPVSU5}%
\end{equation}
which collapses to $%
\bm{\lambda}%
$, $\boldsymbol{\lambda}  ^{\prime}$, and $\boldsymbol{\lambda}
^{\prime\prime}$ at low-energy. With thus both $\Delta\mathcal{B}$ and
$\Delta\mathcal{L}$ couplings of similar sizes, and no suppression from the
GUT scale since these superpotential couplings are renormalizable, the only
way to prevent a too fast proton decay is to require $\boldsymbol{\lambda}
_{5}^{IJK}$ to be tiny~\cite{SU5RPV}. In $SO(10)$, the situation is a bit
different because the $\mathcal{B}-\mathcal{L}$ violating interactions are
forbidden by the gauge symmetry. Though not automatic, R-parity can arise
unscathed from the spontaneous symmetry breaking chain down to the SM gauge
group~\cite{Mohapatra}.
\end{enumerate}

Resolving all these issues would require a thorough and dedicated study of MFV
in a GUT setting, and is clearly beyond our scope. As said, the goal here is
to illustrate what happens when the leptonic and hadronic flavor spaces are
connected, and point out where MFV could retain some usefulness in connection
with the stability of the proton. So, we concentrate exclusively on the fourth
issue above.

\paragraph{With three generations,}

RPV terms may be immediate to construct in terms of the Yukawa couplings. For
instance, in $SU(5)$, we can write the coupling of Eq.~(\ref{RPVSU5}) as%
\begin{equation}
\boldsymbol{\lambda}  _{5}^{IJK}=\varepsilon^{IJL}(\mathbf{Y}_{5})^{LK}\;.
\label{GUT1}%
\end{equation}
Since $\mathbf{Y}_{5}$ is related to $\mathbf{Y}_{d,e}$,
$\boldsymbol{\lambda}  _{5}^{IJK}$ is not particularly suppressed, and the
induced RPV couplings at the MSSM scale are way too large. In this case, the
MFV construction fails to prevent a too fast proton decay. This can be traced
back to the reduced flavor group, which imposes the coherences $SU(3)_{Q}%
\otimes SU(3)_{U}\otimes SU(3)_{E}\rightarrow SU(3)_{10}$ and $SU(3)_{L}%
\otimes SU(3)_{D}\rightarrow SU(3)_{5}$. Indeed, when $L$ and $D$ (or $Q$,
$U$, and $E$) are no longer allowed to transform independently under the
flavor group, a coupling like $\boldsymbol{\lambda}  ^{\prime IJK}L^{I}%
Q^{J}D^{K}$ can be constructed using only the Yukawa couplings, e.g. as
$\varepsilon^{IJK}L^{I}(\mathbf{Y}_{d}Q)^{J}D^{K}$ (compare with
Eq.~(\ref{GUT1})). The need for a neutrino Majorana mass is circumvented, so
the $\Delta\mathcal{L}=\pm1$ and $\Delta\mathcal{B}=\pm1$ couplings are
equally large. The lifetime of the proton cannot be naturally explained by the
flavor structure of the $\boldsymbol{\lambda}  _{5}^{IJK}$ coupling, and some
$U(1)$s have to be imposed.

On the contrary, in $SO(10)$, a restriction of the allowed flavor structures
can effectively enforce R-parity. Indeed, if fermion masses derive only from
Yukawa couplings to Higgses in the $\mathbf{10}$ and $\mathbf{126}$
representations, the only available spurions transform as $\mathbf{6}$s under
$SU(3)_{16}$, and do not permit to construct superpotential RPV terms. Of
course, this observation is relevant only when the assumed breaking scheme
does not automatically lead to R-parity at low energy~\cite{Mohapatra}, so for
example when $\mathcal{B}-\mathcal{L}$ gets broken by a Higgs in the
$\mathbf{16}$ representation.

From these two examples, it is clear that MFV is far less powerful in a GUT
setting, but could still prove useful in some scenarios. So, the main message
is that one should keep an eye on the transformation properties of the various
flavor-symmetry breaking terms, as they could hold the key to proton stability.

\paragraph{With four generations,}

there is no way to form flavor-symmetric combinations of an odd number of
matter superfields since the spurions and the $SU(4)$ invariant tensors all
have an even number of indices. For example, in $SU(5)$, the minimal spurion
content forbids the coupling of Eq.~(\ref{RPVSU5}),%
\begin{equation}
\boldsymbol{\lambda}  _{5}^{IJK}=0\;. \label{GUT2}%
\end{equation}
The same holds in $SO(10)$, for both renormalizable and nonrenormalizable
couplings inducing a low-energy R-parity violation. The only way to violate
Eq.~(\ref{GUT2}) is to allow for a spurion with an odd number of flavor
indices. But with the minimal matter content, this spurion would necessarily
violate R-parity, so by assumptions, it is forbidden. Said differently, the
couplings inducing RPV effects and those inducing the SM Yukawa couplings are
entirely decoupled when there are four generations. They transform in
radically different ways under the unified flavor group. So, the latter cannot
be used to set a natural scale for the former.

As discussed before, nonperturbative Yukawa couplings would not invalidate
Eq.~(\ref{GUT2}) since it is simply forbidden. But, let us stress, if for some
other reasons, perturbativity is nevertheless enforced by extending the matter
content, there is no guarantee that Eq.~(\ref{GUT2}) holds. Specifically, if a
coupling with an odd number of flavor indices is needed, it can never be
expressed in terms of the Yukawa couplings, hence must be part of the spurion
content. But then, it would immediately set a nontrivial natural scale for
$\Delta\mathcal{B}$ and $\Delta\mathcal{L}$ couplings. Given the tight
experimental bounds, either this new spurion is particularly fine-tuned, or
R-parity has to be enforced by some other means.

\section{Conclusion}

In the present paper, the flavor structures of the baryon and lepton number
violating couplings have been systematically analyzed. The main technique was
to relate their flavor structures to those needed to account for the quark and
lepton masses and mixings, using the flavor-symmetry breaking language of
minimal flavor violation. Our results can be split into those relevant for
three and four generations:

\paragraph{With three generations,}

the minimal flavor structures needed to account for the SM fermion masses and
mixings are sufficient to set a natural scale for all the effective
$\Delta\mathcal{B}$ or $\Delta\mathcal{L}$ interactions invariant under
$SU(3)_{C}\otimes SU(2)_{L}\otimes U(1)_{Y}$ (see Table~\ref{TableCCL}). The
Yukawa couplings $\mathbf{Y}_{u}$, $\mathbf{Y}_{d}$, and $\mathbf{Y}_{e}$
permit to construct $\Delta\mathcal{B}=\pm3/N_{c}$ and $\Delta\mathcal{L}%
=\pm3$ effective operators (with $N_{c}=3$ the number of QCD colors), while
$\Delta\mathcal{L}=\pm1$ operators must be proportional to the neutrino
Majorana mass. In some sense, this implements a picture where $\Delta
\mathcal{L}=\pm1$ effects occur concurrently with the seesaw mechanism.
Numerically, because of the smallness of the neutrino masses, and of the large
hierarchies present in the Yukawa couplings, the Wilson coefficients of all
the operators inducing proton decay are sufficiently suppressed to pass the
tight experimental bounds (see the Appendix for an analysis of the
dimension-six Weinberg operators~(\ref{seesaw3})), even for TeV-scale new
physics, provided of course this new physics does not introduce any new flavor structure.

\begin{table}[t]
\centering
\begin{tabular}
[c]{lccccc}\hline
& Spurions & \multicolumn{2}{c}{Three flavors} & \multicolumn{2}{c}{Four
flavors}\\\cline{3-6}
&  & Dim.\  & $\pm(\Delta\mathcal{B},\Delta\mathcal{L})$ & \ Dim.\  &
$\pm(\Delta\mathcal{B},\Delta\mathcal{L})$\\\hline
SM gauge & -- & $27$ & $\mathbf{(6,0)}$, $\mathbf{(0,6)}$, $\mathbf{(3,9)}$,
$\mathbf{(3,\pm3)}$ & $24$ & $\mathbf{(4,4)}$\\
&  & $18$ & $\mathbf{(3,3)}$ & $18$ & $\mathbf{(4,0),(0,4)}$\\
Higgses & $\mathbf{Y}_{u,d,e}$ & $9$ & $\mathbf{(1,3),(2,0)}$ & $10$ &
$\mathbf{(0,4)}$\\
Seesaw & $\;\mathbf{Y}_{u,d,e},\mathbf{\Upsilon}_{\nu}\;$ & $6$ & $(1,1)$ &
$7$ & $(0,2)$\\\hline
MSSM & $\mathbf{Y}_{u,d,e},\mathbf{\Upsilon}_{\nu}$ & $4$ & $(1,1)$ & $5$ &
$\mathbf{(0,4)}$\\
&  & $3$ & $\mathbf{(1,0)},(0,1)$ & $4$ & $(0,2)$\\
SUSY-GUT\  & $\mathbf{Y}_{10,\bar{5}}$ & $3$ & $\mathbf{(1,0),(0,1)}$ & $3$ &
$-$\\\hline
\end{tabular}
\caption{Leading operators violating baryon ($\mathcal{B}$) or lepton
($\mathcal{L}$) number. The second column indicates the allowed spurion
content in each case (see main text for more details). The operators not
suppressed by the neutrino masses $\boldsymbol{\Upsilon}  _{\nu}$ are
indicated in bold. The dimensions refer to the Lagrangian terms for the first
three scenarios, and to the superpotential terms for the last two. For
SUSY-GUT, we take $SU(5)$ for definiteness, and consider only renormalizable
couplings since the usual $\Delta\mathcal{B}$, $\Delta\mathcal{L}$ effective
interactions suppressed by the GUT scale can always be constructed.}%
\label{TableCCL}%
\end{table}

\paragraph{With four generations,}

the Yukawa couplings $\mathbf{Y}_{u}$, $\mathbf{Y}_{d}$, and $\mathbf{Y}_{e}$
permit to construct $\Delta\mathcal{B}=\pm4/N_{c}$ and $\Delta\mathcal{L}=\pm4$ effective operators ($\Delta\mathcal{L}=\pm2$ with a Majorana neutrino mass term), but the former are not color singlets when $N_{c}=3$. Imposing the SM gauge invariance forces the introduction of at least twelve quark fields to violate $\mathcal{B}$ (see Table~\ref{TableCCL}), since $n=3$ is the smallest integer for which $\Delta\mathcal{B}=\pm4n/3$ is also an integer. Thus, the minimal spurion content cannot be used to set a natural scale for the effective $\Delta\mathcal{B}$ and $\Delta\mathcal{L}$ interactions inducing proton decay. For example, the flavor-symmetry properties of the Wilson coefficients of the dimension-six Weinberg operators~(\ref{seesaw3}) are orthogonal to that of the Yukawa couplings. So, if new physics does not introduce drastically new flavor structures, these operators are forbidden, along with any other $\Delta\mathcal{B}=\pm1$ operator, and the proton is absolutely stable. Translated in the four-generation MSSM context, the minimal spurion content effectively enforces not only R-parity, but also the near conservation of $\mathcal{B}$, so that proton decay and neutron oscillations are forbidden, and the lightest supersymmetric particle is stable. In this sense, the present work provides a motivation for the introduction of a sequential fourth generation, besides the replacement of R-parity by MFV. Together with the current clues from the electroweak and flavor sectors~\cite{4GFlavor,4GEWO}, this most simple extension of the SM (or MSSM) appears very appealing.\medskip

In summary, the key to proton stability could well be hidden in a highly
nongeneric flavor structure for the $\Delta\mathcal{B}$ or $\Delta
\mathcal{L}$ interactions. With three generations, these couplings are no more
(or less) fine-tuned than the known fermion masses and mixings, while with
four generations, their drastically different flavor properties rules them out
in a natural way. So, the accidental conservation of $U(1)_{\mathcal{B}}$ and
$U(1)_{\mathcal{L}}$ by the SM Lagrangian seems to be indeed fortuitous, while
imposing R-parity on the MSSM appears redundant. Evidently, what is missing in
this picture is a dynamical mechanism to enforce the minimality of the MFV
prescription at low-energy. The phenomenological successes of this
prescription, both accounting for the suppression of proton decay and the
absence of new physics effects in $K$ and $B$ physics, should act as clear
incentives to pursue this goal.

\subsection*{Acknowledgements}

Many thanks to J.-M. G\'{e}rard, R. Zwicky, and S. Davidson for the
interesting discussions.\pagebreak 

\appendix

\section{Numerical estimates for the proton decay rate}

The operators of Eq.~(\ref{seesaw3}) can induce proton decay. When the spurion
content is limited to $\mathbf{\Upsilon}_{\nu}$ and $\mathbf{Y}_{u,d,e}$,
their Wilson coefficients are predicted, up to $\mathcal{O}(1)$ factors. The
goal of this Appendix is to detail this prediction, and compare it with the
current bounds on the proton lifetime. We do not intend to perform a complete
analysis and get a numerical estimate for the Wilson coefficient of each
possible operator. Rather, our goal is to get the lowest scale $\Lambda$
compatible with proton decay constraints, assuming it is the same for all the
operators in Eq.~(\ref{seesaw3}).

Since there are many ways to insert spurions and contract the flavor indices,
it may not appear immediately obvious looking at Eq.~(\ref{seesaw3}) which
operator and flavor structure is most effective at inducing proton decay. To
identify this operator, first note that those involving the $E$ field will not
be competitive compared to those with $L$, since they necessitate a
$\mathbf{Y}_{e}$ insertion to connect to the $SU(3)_{L}$ space where
$\mathbf{\Upsilon}_{\nu}$ lives. For example, a contraction like%
\begin{equation}
\varepsilon^{IJK}(\mathbf{\Upsilon}_{\nu}\mathbf{Y}_{e}^{\dagger}%
\mathbf{Y}_{e})^{IJ}(E\mathbf{Y}_{e})^{K}\;,
\end{equation}
costs at best a tiny factor $m_{\mu}/v_{d}$ since $E$ cannot be a $\tau
_{R}^{\dagger}$ and $\mathbf{Y}_{e}=\mathbf{m}_{e}/v_{d}$ is diagonal. We thus
remain with $c_{1}^{IJKL}L^{I}Q^{J}Q^{K}Q^{L}$ and $c_{2}^{IJKL}L^{I}%
Q^{J}D^{\dagger K}U^{\dagger L}$. Using either the epsilon tensor of
$SU(3)_{Q}$, $SU(3)_{U}$, or $SU(3)_{D}$, the flavor contractions with the
least number of Yukawa insertions are:%
\begin{align}
c_{1}^{IJKL}L^{I}Q^{J}Q^{K}Q^{L} &  =a_{1}\varepsilon^{IMN}(\mathbf{\Upsilon
}_{\nu}\mathbf{Y}_{e}^{\dagger}\mathbf{Y}_{e})^{MN}L^{I}\times\varepsilon
^{JKL}Q^{J}Q^{K}Q^{L}\nonumber\\
&  \;\;+a_{2}\varepsilon^{IMN}(\mathbf{\Upsilon}_{\nu}\mathbf{Y}_{e}^{\dagger
}\mathbf{Y}_{e})^{MN}L^{I}\times\varepsilon^{JKL}(\mathbf{Y}_{u}%
Q)^{J}(\mathbf{Y}_{u}Q)^{K}(\mathbf{Y}_{u}Q)^{L}\nonumber\\
&  \;\;+a_{3}\varepsilon^{IMN}(\mathbf{\Upsilon}_{\nu}\mathbf{Y}_{e}^{\dagger
}\mathbf{Y}_{e})^{MN}L^{I}\times\varepsilon^{JKL}(\mathbf{Y}_{d}%
Q)^{J}(\mathbf{Y}_{d}Q)^{K}(\mathbf{Y}_{d}Q)^{L}\;,\\
c_{2}^{IJKL}L^{I}Q^{J}D^{\dagger K}U^{\dagger L} &  =a_{4}\varepsilon
^{IMN}(\mathbf{\Upsilon}_{\nu}\mathbf{Y}_{e}^{\dagger}\mathbf{Y}_{e}%
)^{MN}L^{I}\times\varepsilon^{JKL}Q^{J}(D\mathbf{Y}_{d})^{\dagger
K}(U\mathbf{Y}_{u})^{\dagger L}\nonumber\\
&  \;\;+a_{5}\varepsilon^{IMN}(\mathbf{\Upsilon}_{\nu}\mathbf{Y}_{e}^{\dagger
}\mathbf{Y}_{e})^{MN}L^{I}\times\varepsilon^{JKL}(\mathbf{Y}_{u}%
Q)^{J}(D\mathbf{Y}_{d}\mathbf{Y}_{u}^{\dagger})^{\dagger K}U^{\dagger
L}\nonumber\\
&  \;\;+a_{6}\varepsilon^{IMN}(\mathbf{\Upsilon}_{\nu}\mathbf{Y}_{e}^{\dagger
}\mathbf{Y}_{e})^{MN}L^{I}\times\varepsilon^{JKL}(\mathbf{Y}_{d}%
Q)^{J}D^{\dagger K}(U\mathbf{Y}_{u}\mathbf{Y}_{d}^{\dagger})^{\dagger
L}\;,\label{C2IJKL}%
\end{align}
with $a_{i}$ some $\mathcal{O}(1)$ numerical constants. The $L^{I}%
Q^{J}D^{\dagger K}U^{\dagger L}$ operator is systematically suppressed by
light quark mass factors because it involves the quark $SU(2)_{L}$ singlets.
Indeed, to contribute to proton decay, $U$ must be an up quark and $D$ either
a down or strange quark. Since $\mathbf{Y}_{u}=\mathbf{m}_{u}V/v_{u}$ and
$\mathbf{Y}_{d}=\mathbf{m}_{d}/v_{d}$ (see Eq.~(\ref{background})), the
$a_{4}$, $a_{5}$, and $a_{6}$ terms of $c_{2}^{IJKL}$ are all suppressed by
the tiny $m_{u}/v_{u}$ and/or $m_{d,s}/v_{d}$. Such suppressions clearly
remain if more Yukawa spurions are inserted, since all the entries of
$\mathbf{Y}_{u}^{\dagger}\mathbf{Y}_{u}$ and $\mathbf{Y}_{d}^{\dagger
}\mathbf{Y}_{d}$ are smaller than one.

By contrast, the $LQ^{3}$ operator involves quark fields of the same type, so
they can be immediately contracted into a flavor singlet. One should further
keep in mind that the up-quark components of $Q$ are not mass eigenstates (see
Eq.~(\ref{background})), so $Q^{2,3}$ do contain the up-quark. The main
subtlety is that the $a_{1}$ term of $c_{1}^{IJKL}$ actually vanishes because
$L^{I}Q^{J}Q^{K}Q^{L}$ is symmetric under $L\leftrightarrow K$, as can be seen
writing down the $SU(3)_{C}$ and $SU(2)_{L}$ antisymmetric contractions
explicitly, so we replace it by%
\begin{equation}
a_{1}^{\prime}\varepsilon^{IMN}(\mathbf{\Upsilon}_{\nu}\mathbf{Y}_{e}%
^{\dagger}\mathbf{Y}_{e})^{MN}L^{I}\times\varepsilon^{JKO}(\mathbf{Y}%
_{u}^{\dagger}\mathbf{Y}_{u})^{OL}Q^{J}Q^{K}Q^{L}\;.
\end{equation}
Let us concentrate on this term. As all the entries of $\mathbf{Y}%
_{u}^{\dagger}\mathbf{Y}_{u}$ but $(\mathbf{Y}_{u}^{\dagger}\mathbf{Y}%
_{u})^{33}\approx0.5$ are small, we set $L,O=3$ and extract from the
$\bar{\ell}_{L}^{C}u_{L}^{\prime}\bar{s}_{L}^{C}t_{L}^{\prime}$ and $\bar{\nu
}_{L}^{C}d_{L}\bar{s}_{L}^{C}t_{L}^{\prime}$ operators a piece involving only
light quarks using $t_{L}^{\prime}=V_{td}^{\dagger}u_{L}$. With $V_{td}%
\sim10^{-2}$, this suppression is slightly less expensive than using the
off-diagonal entry $(\mathbf{Y}_{u}^{\dagger}\mathbf{Y}_{u})^{31}$,
proportional to $V_{ub}\sim10^{-3}$. A similar suppression can be derived from
the $a_{2}$ contraction, while the $a_{3}$ term is necessarily smaller since
$\mathbf{Y}_{d}$ is diagonal. By the way, this also shows that the dominant
proton decay modes are into strange mesons. Those into nonstrange mesons are
induced by $\ell_{L}c_{L}^{\prime}d_{L}t_{L}^{\prime}$, $c_{L}^{\prime}%
=V_{cd}^{\dagger}u_{L}$, and are thus Cabibbo-suppressed.

On the lepton side, the suppression brought in by the neutrino mass spurion is
much more important. Setting the reactor and atmospheric mixing angles to
$\theta_{13}=0%
{{}^\circ}%
$ and $\theta_{atm}=45%
{{}^\circ}%
$ \cite{PDG}, the background value for the $\mathbf{\Upsilon}_{\nu}$ spurion
can be written as%
\begin{equation}
\mathbf{\Upsilon}_{\nu}\approx\frac{1}{v_{u}}\left(  m_{\nu}\mathbf{1}%
+\frac{\Delta m_{21}e^{-2i\alpha}}{\sqrt{2}\left(  1+t_{\odot}^{2}\right)
}\left(
\begin{array}
[c]{ccc}%
\sqrt{2}t_{\odot}^{2} & t_{\odot} & -t_{\odot}\\
t_{\odot} & 1/\sqrt{2} & -1/\sqrt{2}\\
-t_{\odot} & -1/\sqrt{2} & 1/\sqrt{2}%
\end{array}
\right)  +\frac{\Delta m_{31}e^{-2i\beta}}{2}\left(
\begin{array}
[c]{ccc}%
0 & 0 & 0\\
0 & 1 & 1\\
0 & 1 & 1
\end{array}
\right)  \right)  \;, \label{SpurionNu}%
\end{equation}
where $\theta_{\odot}$ is the solar mixing angle (we set $\tan\theta_{\odot
}=2/3$ in the following), $\alpha$ and $\beta$ are the Majorana phases, and
the mass-eigenstates are written in terms of an overall scale as $m_{\nu
1}=m_{\nu}$, $m_{\nu2}=m_{\nu}+\Delta m_{21}$, $m_{\nu3}=m_{\nu}+\Delta
m_{31}$. Depending on the spectrum (i.e., whether $\nu_{1}$ or $\nu_{3}$ is
the lightest neutrino), the mass differences $\Delta m_{21}$ and $\Delta
m_{31}$ are related to the mixing parameters as%
\begin{equation}
\Delta m_{21}=(\Delta m_{\odot}^{2}+m_{\nu}^{2})^{1/2}-m_{\nu}>0,\;\left\{
\begin{array}
[c]{l}%
\Delta m_{31}=(\Delta m_{atm}^{2}+m_{\nu}^{2})^{1/2}-m_{\nu}%
>0\;\;\text{(Normal),}\\
\Delta m_{31}=(m_{\nu}^{2}-\Delta m_{atm}^{2})^{1/2}-m_{\nu}%
<0\;\;\text{(Inverted).}%
\end{array}
\right.  \label{SpurionNu2}%
\end{equation}
Therefore, for fixed $\Delta m_{\odot}^{2}\approx8\cdot10^{-5}$ eV$^{2}$ and
$\Delta m_{atm}^{2}\approx2\cdot10^{-3}$ eV$^{2}$ \cite{PDG}, the off-diagonal
elements of $\mathbf{\Upsilon}_{\nu}$ quickly decrease with increasing
$m_{\nu}$. Discarding the Majorana phases (irrelevant here), and setting
$\tan\beta=1$ (to recover the SM Higgs sector), we find for the normal
hierarchy%
\begin{equation}
\varepsilon^{IJK}(\mathbf{\Upsilon}_{\nu}\mathbf{Y}_{e}^{\dagger}%
\mathbf{Y}_{e})^{IJ}\approx\frac{1}{v^{3}}\left(
\begin{array}
[c]{c}%
\frac{1}{2}m_{\tau}^{2}\Delta m_{31}\\
\frac{1}{3}m_{\tau}^{2}\Delta m_{21}\\
\frac{1}{3}m_{\mu}^{2}\Delta m_{21}%
\end{array}
\right)  \approx\left(
\begin{array}
[c]{c}%
10^{-17}\rightarrow10^{-19}\\
10^{-18}\rightarrow10^{-21}\\
10^{-21}\rightarrow10^{-23}%
\end{array}
\right)  \text{ for }m_{\nu}=0\rightarrow1\text{ eV}\;. \label{MajoranaDB}%
\end{equation}
Since $m_{\nu}$ can vary only between $(\Delta m_{atm}^{2})^{1/2}$ and about
$1$ eV for the inverted hierarchy, smaller values are found. It is important
to note that only neutrino mass differences appear, as the first term of
$\mathbf{\Upsilon}_{\nu}$ gets projected out in the antisymmetric contraction.
That is why the suppression is maximal for heavy neutrinos. Another
consequence is that $\varepsilon^{IJK}(\mathbf{\Upsilon}_{\nu}\mathbf{Y}%
_{e}^{\dagger}\mathbf{Y}_{e})^{IJ}$ has an inverted hierarchy compared to
$\mathbf{Y}_{e}$, having its first-generation entry significantly larger. In
other words, the dominant proton decay channels are $p^{+}\rightarrow
e^{+}K^{0}$ and $p^{+}\rightarrow\bar{\nu}_{e}K^{+}$, as induced by
$\bar{e}_{L}^{C}u_{L}\bar{s}_{L}^{C}u_{L}$ and $\bar{\nu}_{eL}^{C}d_{L}\bar
{s}_{L}^{C}u_{L}$ (strictly speaking, the $\nu_{e}$ is not a mass eigenstate,
but this is of no concern since it is not detected).

The final MFV prediction is thus $c_{1}^{eusu}\sim10^{-19}\rightarrow10^{-21}$
for $m_{\nu}\sim0\rightarrow1$ eV for $a_{1}\sim1$. From this, we can estimate
the proton decay rate as%
\begin{equation}
\Gamma_{p^{+}}\approx\frac{\alpha_{p}^{2}m_{p}}{16\pi F_{K}^{2}}\left(
1-m_{K}^{2}/m_{p}^{2}\right)  ^{2}\frac{\left|  c_{1}^{eusu}\right|  ^{2}%
}{\Lambda^{4}}\approx10^{-65}\times a_{1}^{2}\left(  \frac{c_{1}^{eusu}%
}{10^{-21}}\right)  ^{2}\left(  \frac{0.003\,\text{GeV}^{3}}{\alpha_{p}%
}\right)  ^{2}\left(  \frac{10\,\text{TeV}}{\Lambda}\right)  ^{4}\;,
\end{equation}
to be compared with the current limits, $\Gamma(p^{+}\rightarrow e^{+}%
K^{0})<1.4\cdot10^{-64}$ and $\Gamma(p^{+}\rightarrow\bar{\nu}K^{+}%
)<3.1\cdot10^{-65}$. For the Cabibbo-suppressed mode, the current limit
$\Gamma(p^{+}\rightarrow e^{+}\pi^{0})<2.5\cdot10^{-66}$ is tighter, but leads
to a similar scale $\Lambda$ once accounting for $c_{1}^{eudu}/c_{1}%
^{eusu}\sim V_{cd}$. Note that coincidentally, the dominant decay modes are
precisely those for which the experimental constraints are the tightest, so if
the scale $\Lambda$ is high enough for them, the bounds for all the other
modes are easily satisfied.

The conclusion of this numerical analysis is that the $\Delta(\mathcal{B+L})$
scale $\Lambda$ could be as low as about 10 TeV, to be compared to
$\Lambda\gtrsim10^{11}$ TeV when $c_{1}\sim1$. Importantly, this should be
understood as a rough estimate of the true NP scale. First, trivially,
$a_{1}\sim0.1\rightarrow10$ would still qualify as natural, and this
translates as a factor 1/3 to 3 for $\Lambda$. Second, current evaluations for
the relevant hadronic matrix element, parametrized by $\alpha_{p}$, range from
$0.003\ $to $0.03$ GeV$^{3}$, see Ref.~\cite{AokiDNS06}, so the high-end of
this range would require $\Lambda$ to be about three times larger. Third, and
more importantly, this estimate is purely based on the flavor symmetry and
thus misses completely any dynamical effect. The $LQ^{3}$ effective operator
could be suppressed by some gauge and/or loop factors. For example, $\Lambda$
is brought down to about $1$ TeV if $a_{1}\sim\alpha/4\pi$. Further, specific
NP models need not generate all the effective operators with equal weight. For
example, as said in the text, $LQ^{3}$ does not arise at tree-level in the
R-parity violating MSSM. Since all the other operators are significantly more
suppressed by light-quark mass factors, their scale $\Lambda$ can actually be
as low as a few hundred GeV without violating proton decay bounds.

\end{document}